# Revealing the dark side of a bright exciton polariton condensate

J.-M. Ménard[1*], C. Poellmann[1], M. Porer[1], U. Leierseder[1]

E. Galopin[2], A. Lemaître[2], A. Amo[2], J. Bloch[2] and R. Huber[1‡]

1. Department of Physics, University of Regensburg, 93040 Regensburg, Germany

2. CNRS-Laboratoire de Photonique et Nanostructures, Route de Nozay, 91460 Marcoussis, France

[*]jean.michel-menard@physik.uni-regensburg.de
[‡]rupert.huber@physik.uni-regensburg.de

**Condensation of bosons causes spectacular phenomena such as superfluidity or superconductivity. Understanding the nature of the condensed particles is crucial for active control of such quantum phases. Fascinating possibilities emerge from condensates of light-matter coupled excitations, such as exciton polaritons, photons hybridized with hydrogen-like bound electron-hole pairs. So far, only the photon component has been resolved, while even the mere existence of excitons in the condensed regime has been challenged. Here we trace the matter component of polariton condensates by monitoring intra-excitonic terahertz transitions. We study how a reservoir of optically dark excitons forms and feeds the degenerate state. Unlike atomic gases, the atom-like transition in excitons is dramatically renormalized upon macroscopic ground state population. Our results establish fundamental differences between polariton condensation and photon lasing and open possibilities for coherent control of condensates.**

In atomic gases, Bose-Einstein condensation (BEC)[1] relies on radiative transitions between orbital states which have been exploited for laser cooling or to control atom-atom interactions (via the so-called Feshbach resonance mechanism)[2]. In contrast, internal degrees of freedom of quantum fluids formed by elementary excitations in solids[3-7] have remained largely elusive, even though, important information on the nature of the condensing quasiparticles may be encoded in their internal structure.

The situation is particularly striking for the prominent example of exciton polaritons – design-cut bosonic quasiparticles that result from strong light-matter coupling in semiconductor microcavities[4,7-18] (Fig. 1a). When a sufficiently high-quality microcavity is nearly resonant with an interband exciton transition, trapped photons may be emitted and reabsorbed multiple times by the exciton before they are lost by dissipation or cavity leakage. Under such conditions, cavity polaritons emerge as light-matter mixed eigenstates[8]. Normal mode splitting leads to a characteristic anticrossing of the lower (LPB) and upper polariton branches (UPB) as a function of the in-plane momentum $k_\parallel$ (Fig. 1b). Near $k_\parallel = 0$, the effective mass of the LPB may be four orders of magnitude lighter than the bare exciton mass[7]. Furthermore the excitonic admixture that determines the mutual interaction strength of polaritons is adjustable by the frequency detuning $\delta$ of the cavity with respect to the exciton resonance[10]. These properties define the metastable zero-momentum state of the LPB as a unique system to achieve a dynamical BEC.

In fact, a dramatic increase of temporal and spatial photoluminescence (PL) coherence occurring at high densities has been interpreted as a spontaneous macroscopic occupation of the LPB at $k_\parallel = 0$.[4] Related phenomena such as nonlinear Josephson oscillations, vortices and long range order have also been reported on the basis of PL data[12-14,18]. However, PL monitors only the collapse of polaritons into their photonic part, whereas a complementary direct handle on the polariton's material component, the exciton, has been missing. Hence important questions about the very nature of the solid-state BEC have been left unanswered: What is the exact role of excitons in the build-up of the macroscopic quantum state? Does the critical condensation density exceed the Mott transition density, above which Coulomb screening effectively terminates the existence of excitons? The debate has been fuelled by a quantitative theory showing that PL measurements alone cannot fully differentiate between signatures induced by excitons or free carriers[16,17,19]. Moreover, the fact that polariton microcavities share their basic design with vertical cavity surface emitting lasers (VCSELs) operating in the weak coupling regime, suggests that the properties of dynamical BEC may be simply caused by stimulated photon emission.

In this article, we simultaneously observe the photon and the matter part of a condensing exciton polariton gas. To this end we combine PL experiments with terahertz (THz) spectroscopy probing the atom-like internal structure. This concept allows us to directly observe how excitons form and feed a polariton condensate occurring below the Mott density. We quantify a strong polariton population at $k_\parallel = 0$ via a new internal absorption line. The experimental idea is illustrated in Fig. 1c. As hydrogen-like pairs of one electron and one hole, excitons feature an orbital structure. Dipole transitions between the 1s and the 2p states are resonant in the THz spectral range and have been used to probe the formation of free excitons at arbitrary centre of mass momenta or to drive nonlinear optics[21-27]. When embedded into a high-quality semiconductor microcavity, strong light-matter coupling modifies the dispersion of the 1s state at small in-plane momenta. In contrast, the dispersion of the optically dark 2p state remains set by the centre of mass motion of the bare exciton (Fig. 1c). Therefore, the 1s-2p resonance depends on $k_\parallel$ and THz spectroscopy may be used to map out the polariton distribution in momentum space.

**Results**

**Photonic component**. Our GaAs/AlGaAs microcavity contains 12 quantum wells located at the antinodes of the optical mode (Supplementary Figure 1 and Supplementary Methods). The sample is kept at a lattice temperature of $T_L = 10$ K and features a small cavity-exciton detuning $\delta = 2$ meV. Normal mode splitting lowers the energy of the LPB at $k_\parallel = 0$ by 8 meV (Fig. 1c). Unbound electron-hole pairs are injected above the band gap by 20-fs optical pulses at a photon energy of $\hbar\omega_p = 1.65$ eV (Supplementary Figures 2 and 3). We identify the onset of polariton condensation by collecting the total PL intensity as a function of the density of photoinjected electron-hole pairs, $\rho$ (Fig. 2a). The superlinear increase of the intensity, above a threshold of $\rho_c = 6 \times 10^{10}$ cm$^{-2}$, is characteristic of what has previously been explained by the emergence of stimulated polariton scattering[4,7]. Angle-resolved PL spectra confirm this conjecture: For a low density $\rho_1 = 5 \times 10^{10}$ cm$^{-2}$ < $\rho_c$ (Fig. 2b), light emission is spread over a broad range of angles -20° < θ < 20°. In contrast, dramatic narrowing of the PL angular distribution is seen above the condensation threshold, at $\rho_2 = 15 \times 10^{10}$ cm$^{-2}$ (Fig. 2c).

**Matter component**. Simultaneously we also trace the THz response of the microcavity as a function of the delay time τ after photoinjection of free electron-hole pairs. Electro-optic sampling resolves both amplitude and phase of the THz waveforms transmitted through the sample. The spot size of the incident THz pulse is two times smaller than the excitation area to ensure a homogeneous probing of the carrier dynamics. These data allow us to retrieve the photoinduced changes of the real parts of both the optical conductivity $\Delta\sigma_1$ and the dielectric function $\Delta\varepsilon_1$ (Supplementary Methods). Different pairs of THz emitter and detector crystals allow us to resolve different regions of interest of the THz window (Supplementary Figure 4). Figure 3 depicts spectra for representative delay times τ below the critical density. For τ = 2 ps, the monotonic decrease (increase) of $\Delta\sigma_1$ ($\Delta\varepsilon_1$) indicates that a Drude-like response of the photoinjected carriers prevails[21,22]. At later delay times, a distinct maximum of $\Delta\sigma_1$ develops at $\hbar\omega = 9$ meV accompanied by a dispersive feature in $\Delta\varepsilon_1$ due to the 1s-2p transition of bare excitons expected in the 7-nm-wide GaAs quantum wells[28]. THz absorption at this energy, thus, measures the exciton population. As seen from the amplitude of the resonance features, the exciton density grows on the time scale of 100 ps before a decay dynamics sets in on the nanosecond scale[21].

Interestingly, the energy position of the exciton line shifts from $\hbar\omega = 9$ meV, at $\tau = 40$ ps, to 11 meV, at $\tau = 1$ ns. We attribute this dynamics to polariton cooling: The quasiparticles are redistributed from large-momentum states – called the reservoir – towards the bottleneck region near the inflection point of the polariton dispersion[7], where the strong non-parabolicity of the LPB blueshifts the internal 1s-2p resonance (Fig. 1c, green arrow). As a first approximation we compare our experimental results with a two-component model of excitons and unbound electron-hole pairs[21,22]. By adjusting the energy of the 1s-2p transition as a free parameter, $\Delta\sigma_1$ and $\Delta\varepsilon_1$ are simultaneously well reproduced (Fig. 3, black curves). A quantitative summary of the densities of free carriers ($\rho_{e-h}$) and the exciton reservoir ($\rho_X$) retrieved from the model fits is given in Fig. 4d: While the free carriers dominate immediately after photoinjection, their density drops below 20% for delay times $\tau \geq 200$ ps. At the same time the exciton reservoir fills up to a maximum density of $\rho_X(\tau = 200 \text{ ps}) = 0.8 \times \rho_1$, followed by a nanosecond decay.

Increasing the pump fluence may ultimately drive a Mott transition in the system. In this situation, excitons completely cease to exist due to strong Coulomb screening, strong coupling is lost and any lasing feature in the PL is related to standard photon lasing under population inversion and not to polariton condensation[29]. Photon lasing in the weak-coupling regime is expected to lead to a Drude-like THz response (Supplementary Figure 5) instead of excitonic resonances.

In order to test whether a Mott transition is already reached when the PL indicates condensation, we record the THz dielectric response at $\rho_2 > \rho_c$ (Fig. 4a). Again the Drude response of free carriers dominates initially ($\tau \leq 10$ ps). While a sizeable fraction of unbound electron-hole pairs now survives even beyond these early times, excitons in the reservoir clearly start to build up. The corresponding peak in $\Delta\sigma_1$ (Fig. 4a) and the dispersion in $\Delta\varepsilon_1$ (Supplementary Figure 6) occur at $\hbar\omega = 9$ meV, for $\tau = 50$ ps, followed by a qualitatively similar blueshift to that seen above (Fig. 3a). The temporal traces of $\rho_{e\text{-}h}$ and $\rho_X$ extracted from the two-component model are depicted in Fig. 4e. While the weight of free carriers is qualitatively different from the situation observed at low fluence, the sheer existence of excitons above the condensation threshold demonstrates that the system, indeed, remains below the Mott transition and strong light-matter coupling regime persists.. A systematic study of the free and bound carrier densities as a function of the pump fluence confirms this conclusion (Supplementary Figure 7).

Most importantly, the THz response also features an additional novel manifestation of the polariton condensate. For 50 ps < $\tau$ < 150 ps, the data significantly depart from the two-component model. In the vicinity of $\hbar\omega = 17$ meV, i.e. far above intra-excitonic resonances in the reservoir, an additional peak in $\Delta\sigma_1$ (Fig. 4a, red shadow) and a concomitant dispersive slope of $\Delta\varepsilon_1$ develops (Supplementary Figure 6). The resonance energy coincides with the expected 1s-2p transition in polaritons located at $k_\parallel = 0$ (Fig. 1c). This excitation is expected to become detectable only when the minimum of the LPB is macroscopically populated. Indeed, the absorption line is not seen for densities $\rho < \rho_c$ while it grows above this threshold (Fig. 4b). Hence we conclude that the additional absorption feature originates from a macroscopic population of the LPB close to $k_\parallel = 0$, indicating the transition of the system into the condensed phase. The frequency of this novel resonance can be consistently shifted by changing $\delta$ (Supplementary Figures 8, 9 and Supplementary Discussion).

Our observation provides qualitatively new proof that the strong coherent emission associated with dynamical condensation is fundamentally different from the operation of VCSELs, which remain in the weak-coupling regime and exploit free carriers. By spectrally integrating the Lorentzian fit (Figs. 4a and b) highlighting the departure of $\Delta\sigma_1$ from the two-component model we obtain a first experimental estimate of the density $\rho_{cond}$ of cold polaritons at $k_\parallel = 0$ as a function of the excitation density (Fig. 4c) and the delay time $\tau$ (Fig. 4e, red spheres). The excitation density required to observe an onset of a macroscopic polariton population at $k_\parallel = 0$ (Fig. 4c) amounts to $\rho = (5 \pm 1) \times 10^{10}$ cm$^{-2}$. Remarkably this value coincides, within error margins, with $\rho_c$, the threshold extracted from the PL data of Fig. 2a. The maximum value of $\rho_{cond}$, reached in our experiment, for $\rho = \rho_2$ and $\tau = 50$ ps (Fig. 4e), amounts to $\rho_{cond} = 6.5 \times 10^9$ cm$^{-2}$, representing 4% of all electron-hole pairs.

Finally, we compare the THz dynamics of reservoir excitons and polaritons at $k_\parallel = 0$ with the complementary time-resolved PL measurements probing the photonic part of polaritons at small momenta. For low density $\rho_1$ (Fig. 4d), the PL intensity trails $\rho_X$ with a distinct delay of more than 50 ps due to multiple scattering events with acoustic phonons that are required to relax the quasiparticles towards small $k_\parallel$ values. Conversely, Fig. 4e shows that the excitonic and photonic components of the cold polariton population follow the same dynamics within the experimental error, indicating that they likely originate from the same macroscopic condensate. A comparison of $\rho_X$ below and above the critical excitation density (Figs. 4d and e) reveals that the decay of the exciton density in the reservoir is accelerated as long as the condensate is present for $\tau < 200$ ps. This feature may be a direct consequence of stimulated bosonic scattering into the macroscopically populated zero-momentum state which may efficiently deplete the exciton reservoir.

Internal degrees of freedom of solid-state quantum fluids provide a fundamentally new access to macroscopic wavefunctions. As a first example, we probe an intra-excitonic transition to map out the matter component of a condensing exciton-polariton gas and explore the real nature of the dynamical condensate. The same concept may be extended to other intriguing condensates based on magnons, photons or excitons[7]. In the future, high-intensity THz pulses may even bring coherent control[24,25,27,30] and optical protocols exploiting the internal degrees of freedom of solid-based quantum fluids into reach.

**References**

1. Anderson, M. H., Ensher, J. R., Matthews, M. R., Wieman, C. E. & Cornell, E. A. Observation of Bose-Einstein Condensation in a Dilute Atomic Vapor. *Science* **269**, 198–201 (1995).
2. Chu, S., Hollberg, L., Bjorkholm, J. E., Cable, A. & Ashkin, A. Three-Dimensional Viscous Confinement and Cooling of Atoms by Resonance Radiation Pressure. *Phys. Rev. Lett.* **55**, 48–51 (1985).
3. Leggett, A. J. Quantum Liquids: Bose Condensation and Cooper Pairing in Condensed-matter Systems (Oxford University Press: Oxford, NY, USA, 2006).
4. Kasprzak J. *et al.* Bose-Einstein condensation of exciton polaritons. *Nature* **443**, 409–414 (2006).
5. Demokritov, S. O. *et al.* Bose-Einstein condensation of quasi-equilibrium magnons at room temperature under pumping. *Nature* **443**, 430–433 (2006).
6. Klaers, J., Schmitt, J., Vewinger, F. & Weitz, M. Bose-Einstein condensation of photons in an optical microcavity. *Nature* **468**, 546–548 (2010).
7. Carusotto, I. & Ciuti, C. Quantum fluids of light. *Rev. Mod. Phys.* **85**, 299–366 (2013) and references therein.
8. Weisbuch, C., Nishioka, M., Ishikawa, A. & Arakawa, Y. Observation of the coupled exciton-photon mode splitting in a semiconductor quantum microcavity. *Phys. Rev. Lett.* **69**, 3314–3317 (1992).
9. Balili, R., Hartwell, V., Snoke, D., Pfeiffer, L. & West, K. Bose-Einstein condensation of microcavity polaritons in a trap. *Science* **316**, 1007–1010 (2007).
10. Kavokin, A., Baumberg, J. J., Malpuech, G. & Laussy, F. P. Microcavities (Oxford University Press: Oxford, NY, USA, 2007).
11. Christopoulos, S. *et al.* Room-temperature polariton lasing in semiconductor microcavities. *Phys. Rev. Lett.* **98**, 126405 (2007).
12. Lagoudakis, K. G. *et al.* Quantized vortices in an exciton–polariton condensate. *Nature Phys.* **4**, 706–710 (2008).
13. Abbarchi, M. *et al.* Macroscopic quantum self-trapping and Josephson oscillations of exciton polaritons. *Nature Phys.* **9**, 275–279 (2013).
14. Wertz, E. *et al.* Spontaneous formation and optical manipulation of extended polariton condensates. *Nature Phys.* **6**, 860–865 (2010).

15. Wiersig, J. *et al.* Direct observation of correlations between individual photon emission events of a microcavity laser. *Nature* **460**, 245–249 (2009).
16. Timofeev, V. & Sanvitto, D. Exciton polaritons in microcavities (Springer: Berlin Heidelberg, Germany, 2012).
17. Gibbs, H. M., Khitrova, G. & Koch, S. W. Exciton–polariton light–semiconductor coupling effects. *Nature Photon.* **5**, 273–282 (2011).
18. Nardin, G. *et al.* Hydrodynamic nucleation of quantized vortex pairs in a polariton quantum fluid. *Nature Phys.* **7**, 635–641 (2011).
19. Chatterjee, S. *et al.* Excitonic photoluminescence in semiconductor quantum wells: plasma versus excitons. *Phys. Rev. Lett.* **92**, 067402 (2004).
20. Kira, M., Hoyer, W., Stroucken, T. & Koch, S. W. Exciton formation in semiconductors and the influence of a photonic environment. *Phys. Rev. Lett.* **87**, 176401 (2001).
21. Kaindl, R. A., Carnahan, M. A., Hägele, D., Lövenich, R. & Chemla, D. S. Ultrafast terahertz probes of transient conducting and insulating phases in an electron-hole gas. *Nature* **423**, 734–738 (2003).
22. Huber, R., Kaindl, R. A., Schmid, B. A. & Chemla, D. S. Broadband terahertz study of excitonic resonances in the high-density regime in GaAs/Al$_x$Ga$_{(1-x)}$As quantum wells. *Phys. Rev. B* **72**, 161314 (2005).
23. Kubouchi, M., Yoshioka, K., Shimano, R., Mysyrowicz, A. & Kuwata-Gonokami, M. Study of Orthoexciton-to-Paraexciton Conversion in Cu$_2$O by Excitonic Lyman Spectroscopy. *Phys. Rev. Lett.* **94**, 016403 (2005).
24. Leinß, S. *et al.* Terahertz coherent control of optically dark paraexcitons in Cu$_2$O. *Phys. Rev. Lett.* **101**, 246401 (2008).
25. Wagner, M. *et al.* Observation of the intraexciton Autler-Townes effect in GaAs/AlGaAs semiconductor quantum wells. *Phys. Rev. Lett.* **105**, 167401 (2010).
26. Zaks, B., Liu, R. B. & Sherwin, M. S. Experimental observation of electron-hole recollisions. *Nature* **483**, 580–583 (2012).
27. Rice, W. D. *et al.* Terahertz-Radiation-Induced Exciton Shelving and Intra-Excitonic Scattering. *arXiv:1203.3994v1*.
28. Bastard, G., Mendez, E. E., Chang, L. L. & Esaki, L. Exciton binding energy in quantum wells. *Phys. Rev. B* **26**, 1974–1979 (1982).
29. Butté, R. *et al.* Transition from strong to weak coupling and the onset of lasing in semiconductor microcavities. *Phys. Rev. B* **65**, 205310 (2002).

30. Tomaino, J. L. *et al.* Terahertz excitation of a coherent Λ-type three-level system of exciton-polariton modes in a quantum-well microcavity. *Phys. Rev. Lett.* **108**, 267402 (2012).

**Acknowledgements**

We thank P. Steinleitner, D. Bougeard, and C. Ciuti for helpful discussions, and I. Gronwald and M. Furthmeier for experimental assistance. Support by the European Research Council via ERC Starting Grant QUANTUMsubCYCLE, the German Research Foundation (DFG) via the Emmy Noether Program (HU1598/1-1), the Alexander von Humboldt Foundation, the French RENATECH network and the Labex NanoSaclay is acknowledged.

**Contributions**

R.H. and J.B. conceived the idea, J.-M.M., C.P., M.P., A.A., J.B. and R.H. designed the experiment. E.G. and A.L. fabricated the sample. J.-M.M., C.P., M.P., U.L. and A.A. conducted the experiment. J.-M.M., C.P., M.P., U.L., A.A., J.B. and R.H. analysed and interpreted the data. J.-M.M., C.P., A.A, J.B. and R.H. prepared the manuscript with the help of all co-authors.

**Competing financial interests**

The authors declare no competing financial interests.

**Figure 1 | Schematic of PL and THz probing of cavity exciton polaritons**. (**a**) Polaritons (pink spheres with blue halo) emerge from strong coupling between the excitonic interband resonance in a quantum well (transparent sheet) and the photonic mode of a GaAs/AlGaAs microcavity. THz probing (blue curve) maps out the matter component of the polaritons while PL (red arrows) leaking through a Bragg mirror reveals the photonic component. (**b**) Normal-mode splitting. The heavy-hole 1s exciton-resonance (dashed curve) and the photonic mode (dotted curve) are replaced by the upper and lower polariton branches (UPB and LPB, respectively; solid curves). PL (thick red arrow) originates from radiative decay of polaritons at small in-plane momenta $k_\parallel$. (**c**) THz absorption probes hydrogen-like intra-excitonic transitions. While the 1s state gets spectrally shifted by strong light-matter coupling, the optically dark 2p exciton is not affected by the cavity. The resulting momentum dependence of the THz transition energy allows us to map out the momentum distribution of the polaritons as they relax towards $k_\parallel = 0$ (green dotted arrow).

**Figure 2 | Probing the photonic part of polaritons at the condensation threshold**. (**a**) Time- and momentum-integrated PL intensity as a function of the density of photoexcited electron-hole pairs $\rho$. Unbound electron-hole pairs are injected by 20-fs pulses centred at $\hbar\omega_p = 1.65$ eV. The steep increase of the PL intensity above a critical density $\rho_c = 6 \times 10^{10}$ cm$^{-2}$ has been assigned to bosonic stimulation due to the formation of a dynamic condensate. Broken lines: guides to the eye underpinning the linear increase of PL intensity in the low- and high-density limits. Angle-resolved PL measurements taken for $\rho_1 = 5 \times 10^{10}$ cm$^{-2}$ (**b**) and $\rho_2 = 15 \times 10^{10}$ cm$^{-2}$ (**c**), i.e. below and above the critical density $\rho_c$. When $\rho$ exceeds $\rho_c$, the broad PL dispersion collapses into a degenerate momentum and energy state indicative of a polariton condensate.

**Figure 3 | Time-resolved THz response of the formation and cooling of reservoir excitons.** (a) Real part of the pump-induced THz conductivity $\Delta\sigma_1$ for various delay times $\tau$ (indicated next to the curves) after photoinjection of unbound electron-hole pairs of a low density of $\rho_1$, by a 20-fs near-infrared pulse ($\hbar\omega_p$ = 1.65 eV). (b) Corresponding real part of the dielectric function, $\Delta\varepsilon_1$. Blue spheres: experimental data, black curves: two-component model simultaneously fitting $\Delta\sigma_1$ and $\Delta\varepsilon_1$. The corresponding densities of excitons ($\rho_X$) and unbound electron-hole pairs ($\rho_{e-h}$) extracted from the fits are displayed in Fig. 4d. ZnTe emitter and detector crystals are used in the experimental setup to resolve the THz signal between 6 and 14 meV.

**Figure 4 | Time-resolved THz response of condensed polaritons.** Real part of the pump-induced THz conductivity $\Delta\sigma_1$ (a) as a function of delay time $\tau$, for $\rho = \rho_2$, and (b) as a function of excitation density $\rho$, for $\tau = 50$ ps. The dielectric function $\Delta\varepsilon_1$ corresponding to both situations is shown in Supplementary Figure 6. Blue spheres: experiment, black broken curves: two-component model fitting simultaneously $\Delta\sigma_1$ and $\Delta\varepsilon_1$. Corresponding densities $\rho_X$ and $\rho_{e-h}$ extracted from the fits to (a) are displayed in (e). Red shadows: A Lorentz function (FWHM = 3 meV) traces the departures between the experimental data and the two-component model due to polariton population at $k_\| = 0$. GaP emitter and detector crystals are used in the experimental setup to resolve the THz signal between 8 and 22 meV. (c) Density $\rho_{cond}$ of polaritons at $k_\| = 0$ as a function of excitation density, as obtained by spectrally integrating the red shaded areas in (b). Broken line: guide to the eye. (d) Densities of free carriers, $\rho_{e-h}$, (green spheres) and reservoir excitons, $\rho_X$, (blue spheres) as extracted by fitting the two-component model to the data of Fig. 3, displayed as a function of $\tau$. The excitation density is $\rho = \rho_1$. The data are compared with the time-resolved PL intensity (black curve) obtained from streak camera measurements (Supplementary Figure 10). (e) Corresponding

data obtained above the critical excitation density $\rho = \rho_2$. $\rho_{\text{e-h}}$ (green spheres), $\rho_X$ (blue spheres), and $\rho_{\text{cond}}$ (red spheres) are extracted from (**a**).

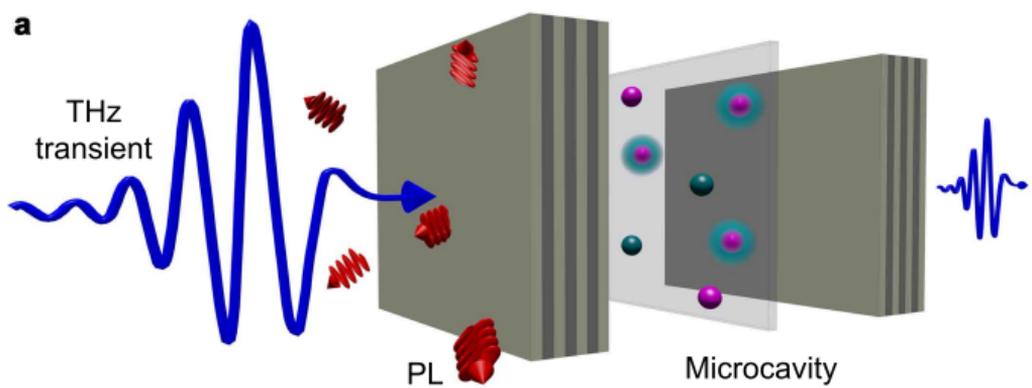

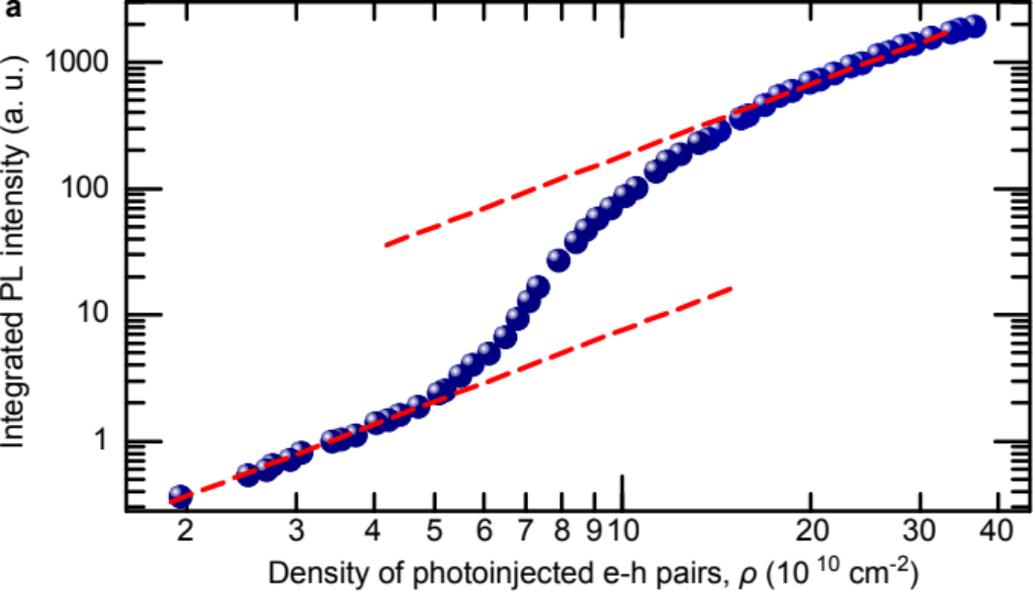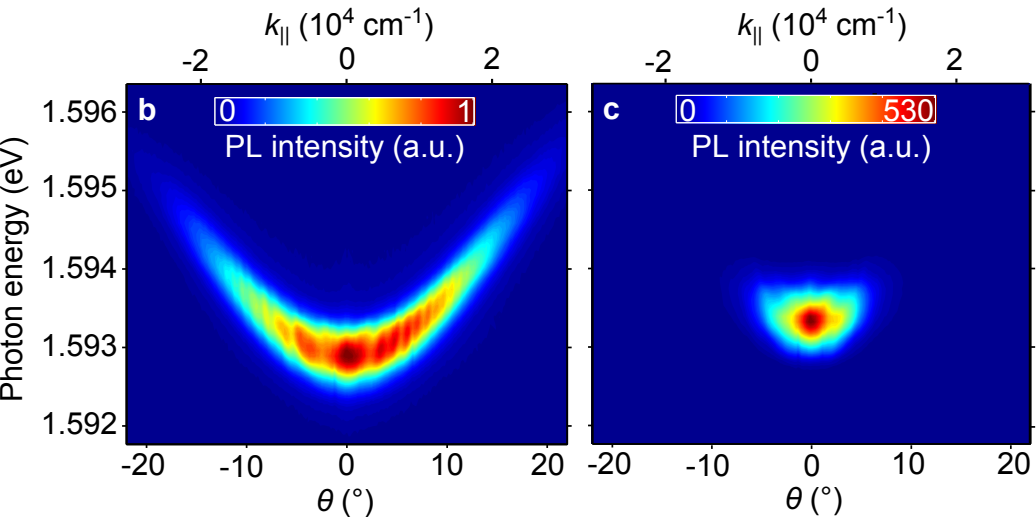

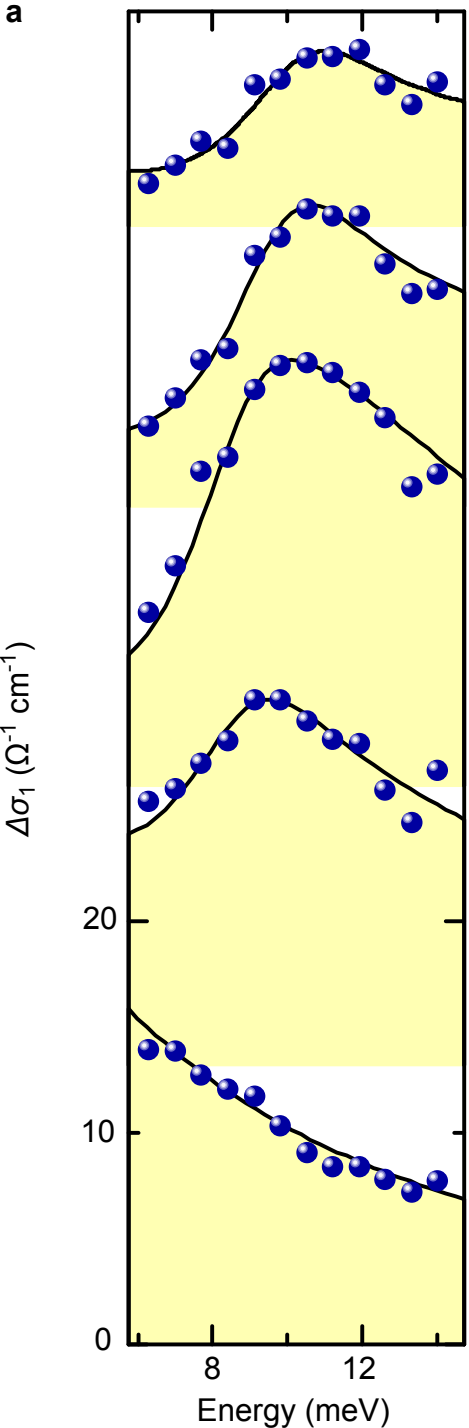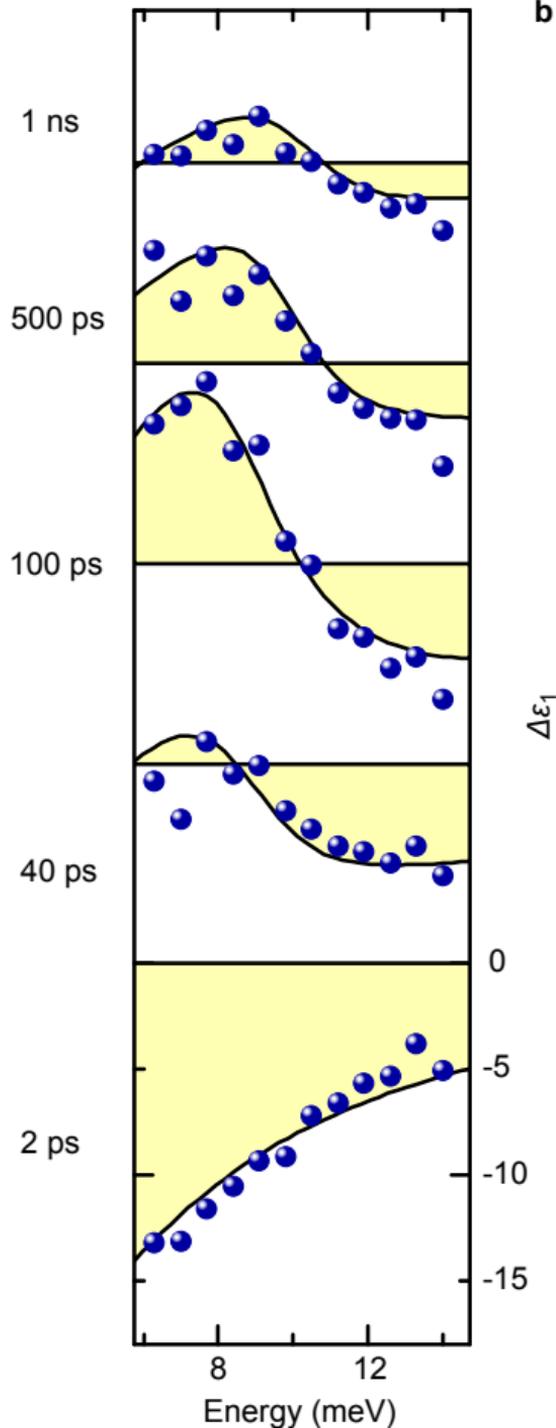

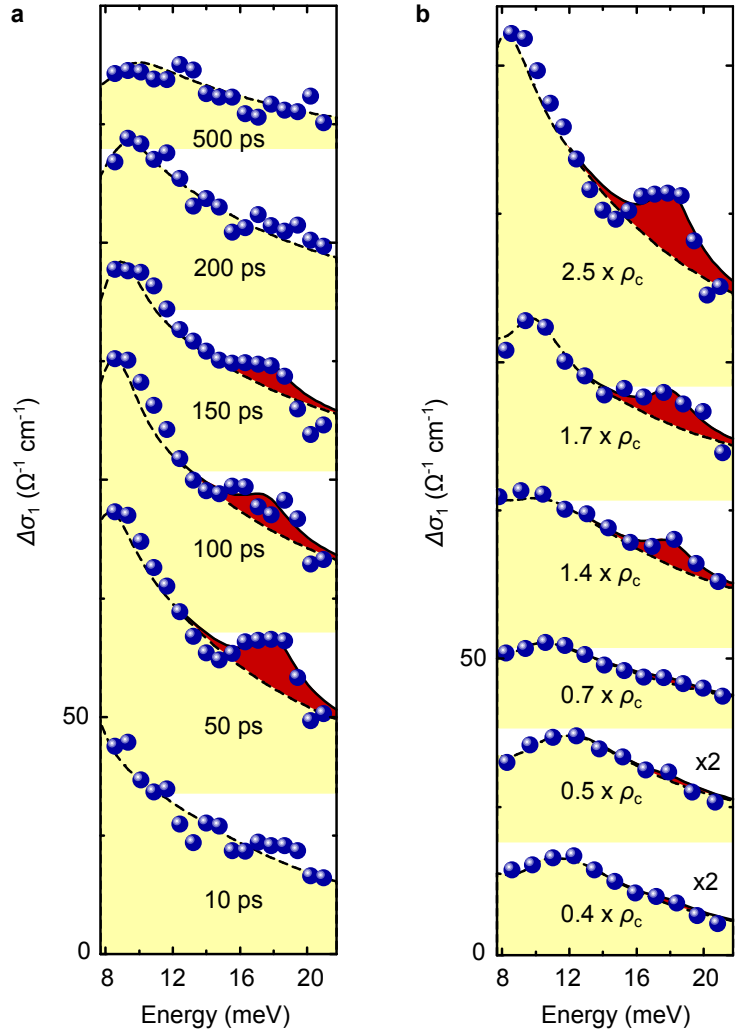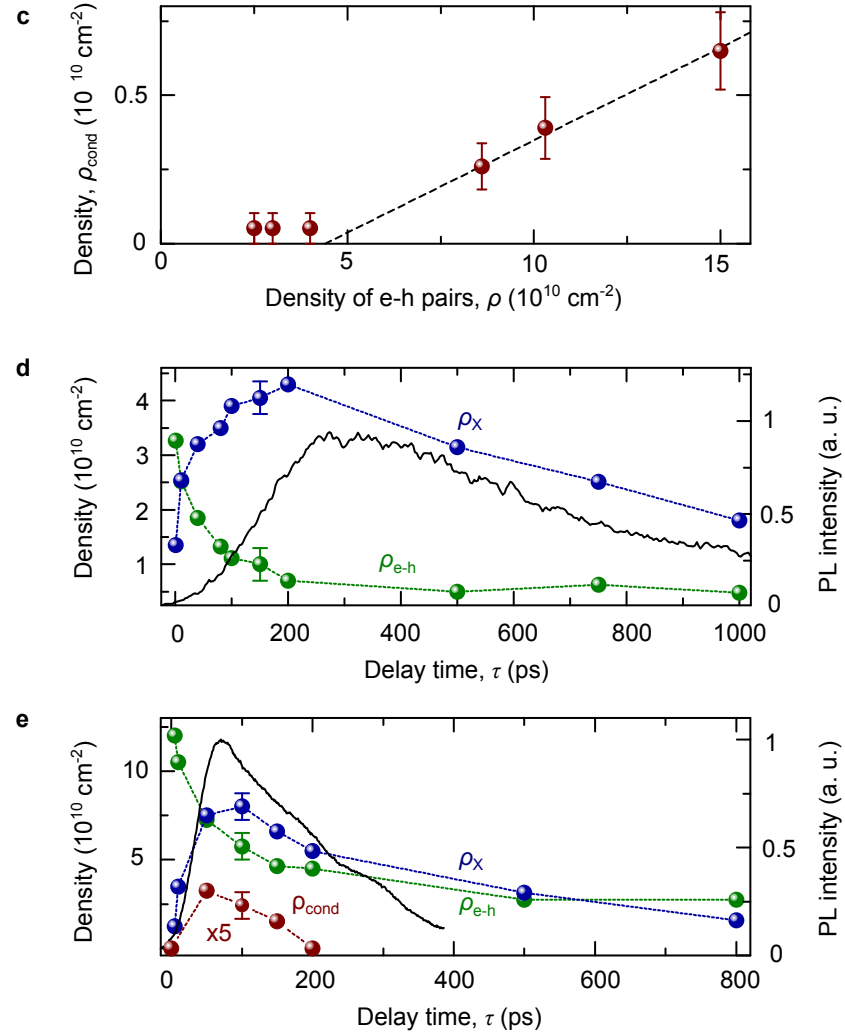